\documentclass{kapproc} 
\usepackage{epsfig}
\begin{document}

\articletitle[]
{The Gamma-ray Properties of \\
Unidentified EGRET Sources}

\author{Olaf Reimer}
\affil{Laboratory for High Energy Astrophysics\\
NASA Goddard Space Flight Center\\ 
Greenbelt, MD 20771\\
USA}
\email{olr@egret.gsfc.nasa.gov}


\begin{abstract}
Although the majority of gamma-ray sources still remain unidentified, we
have various kinds of information to characterize the observational properties
of unidentified EGRET sources. Despite astronomical properties like locations
of individual sources or the collective arrangement of the class as such, the nine
years of CGRO observations provide the ability to investigate flux variability
at different timescales, enable us to perform periodicity searches, determine 
gamma-ray source spectra between 30 MeV and 10 GeV and even investigate 
spectral variability. The basic observational properties of unidentified 
high-energy gamma-ray sources discovered by EGRET are reviewed. Various 
instrumental and observational peculiarities affecting the interpretation 
of the EGRET data are pointed out, also describing the way such biases might 
affect scientific conclusions drawn from the EGRET data.
\end{abstract}

\section{Introduction}
With NASAs Compton Gamma-Ray Observatory mission terminated in June 2000, the 
EGRET data base will remain a unique and extremely important source of scientific 
information. Although being an archival data base from now on, it has to be 
considered as state-of-the-art for several more years, because presently no other 
instrument covering the high-energy gamma-ray wavelengths is in orbit. 
Even when instruments like AGILE and GLAST will become operational, the EGRET data 
will be the reference for new observational results. CGROs coverage of a long time 
period between 1991 and 2000 will be used in determining the long-term 
behaviour of gamma-ray sources and very likely for subsequent archival research. 
To a lesser degree, this has been already demonstrated by comparing EGRET data 
with results from previous missions, in particular with COS-B. Not only were earlier 
reported source detections considered for positional comparison, in cases 
like Geminga archival data were analyzed in conjunction with the EGRET data for 
tracing its long term periodicity behavior. When going into the subject of discussing 
unidentified EGRET sources, COS-B source findings are still an interesting aspect 
of reference, since predictions from COS-B population studies could be supported
or rejected using the EGRET data or to establish the long-term coverage of individual 
sources (i.e. 2CG 135+1 or 2CG 075+0).
Here, I will review how EGRET data were used to construct the source catalogs, 
flux histories and source spectra, and discuss the pecularities of the 
existing point-source catalogs, their positional accuracy and underlying systematics.
The importance of understanding in which way such biases might affect scientific
conclusions will be addressed. Also, different approaches to deal with variability 
are compared. Quantitatively, the flux determination needs to be related to EGRETs 
instrumental response in orbit over time and energy. The determination
of photon spectra will be discribed and a view beyond simple single power-law fits 
needs to be given. In several cases with exceptional observational coverage, also
spectral variabilty could be adressed. 
Having accumulated detailed knowledge of the spacial, temporal and spectral properties 
of individual unidentified EGRET sources, the quest for finding signatures in the 
collective could be challenged. Conclusions are drawn on the validity of assumptions, 
selections and cuts in population studies, mainly under aspects of known instrumental 
biases or pecularities with the gamma-ray point source catalogs and questioning barely 
justified speculations.    

\section{EGRET Source Catalogs and Gamma--Ray Source Locations}

Omitting all low level EGRET data products and therefore the complete process
of event reconstuction and event quality classes, the Third EGRET catalog of 
high-energy gamma-ray sources (Hartman et al. 1999) has been constructed on the 
basis of individual viewing periods, meaningfully selected combinations of individual 
viewing periods (i.e. subsequent short observations or sequences with nearly
identical pointings) and annual superpositions of all observations synchronized
with CGRO observation cycles. For each of those selections
appropriate count, exposure and intensity maps have been constructed, usually 
by applying a standard $<30^\circ$ field--of--view cut. These maps have been 
analysed by means of a maximum-likelihood procedure (Mattox et al. 1996). Although 
fluxes are consistently given for E$> 100$ MeV, similar analyses have been performed
for the energy intervals 300-1000 MeV and above 1000 MeV. The different likelihood test 
statistics (TS) maps were compared and, as long as (TS)$^{1/2}> 4$, the one which produced 
the smallest error contours was chosen to represent the actual source location.
Figure 1 shows the EGRET all-sky TS map, the result of the maximum--likelihood 
analysis procedure carried out from all viewing periods between April 1991 and 
September 1995 at $> 100$ MeV.
\begin{figure}[ht]
\epsfig{file=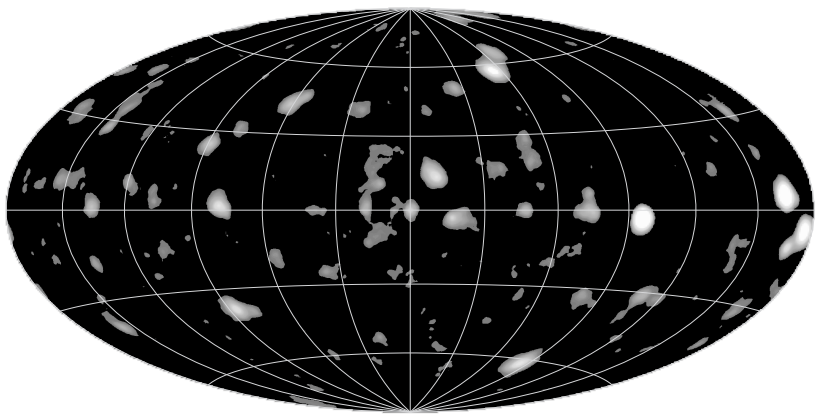, width=0.9\hsize}
\caption{Test statistics (to say detection significance) map as result
of the maximum--likelihood analysis of the EGRET data from cycle 1 to 4
observations}
\end{figure}
The intense and highly structured diffuse emission along the Galactic planes 
makes the determination of sources in the Galactic plane more subject to systematic 
uncertainties than those at higher latitudes. Therefore, the detection criterion 
for a gamma-ray source is also different for its location: In at least one of the 
derived maps (TS)$^{1/2}> 4$ for sources at $|b|>10^\circ$ or (TS)$^{1/2}> 5$ at 
$|b|<10^\circ$ must be fulfilled. This peculiar step in the acceptance criterion 
is somewhat arbitrary, however translates directly into a bias taken care of in source 
population studies. Additionally, the acceptance criterion could be fulfilled in any of
the derived maps, resulting in detections on the basis of very uneven exposure times.
One-time flaring sources will be included as well as sources which build up the 
detection significance in a purely statistical way originating from more and more 
observations, i.e. exposure. The resulting point source catalog (Fig.2) therefore 
represents a rather uneven sample for population study applications.\\
Also, the underlying diffuse emission model (Hunter et al. 1997) is known to be imperfect on
smaller scales. The likelihood source detection algorithm will translate such inaccuracies of the 
diffuse model into the detection significance and, very likely, into acceptance issues 
for sources near the catalog thresholds. The procedure of independently scaling of the
nominal values of the diffuse emission model within the radius--of--analysis (usually $15^\circ$) 
in order to account for small-scale structures might not be in each case the most accurate one 
in order to discriminate point-like excesses against features in the diffuse emission.
In confused regions with sources of significantly overlapping point spread functions (PSF), the 
order in which sources are optimized becomes important, too. These systematics are generally not 
easy to quantify, but obviously need consideration in 3EG catalog-based population studies.
Figure 2 shows the high-energy gamma-ray sources fulfilling the 3EG catalog 
acceptance criteria. The size of the symbols represents the maximum intensity seen 
for this source.\\
\begin{figure}[ht]
\epsfig{file=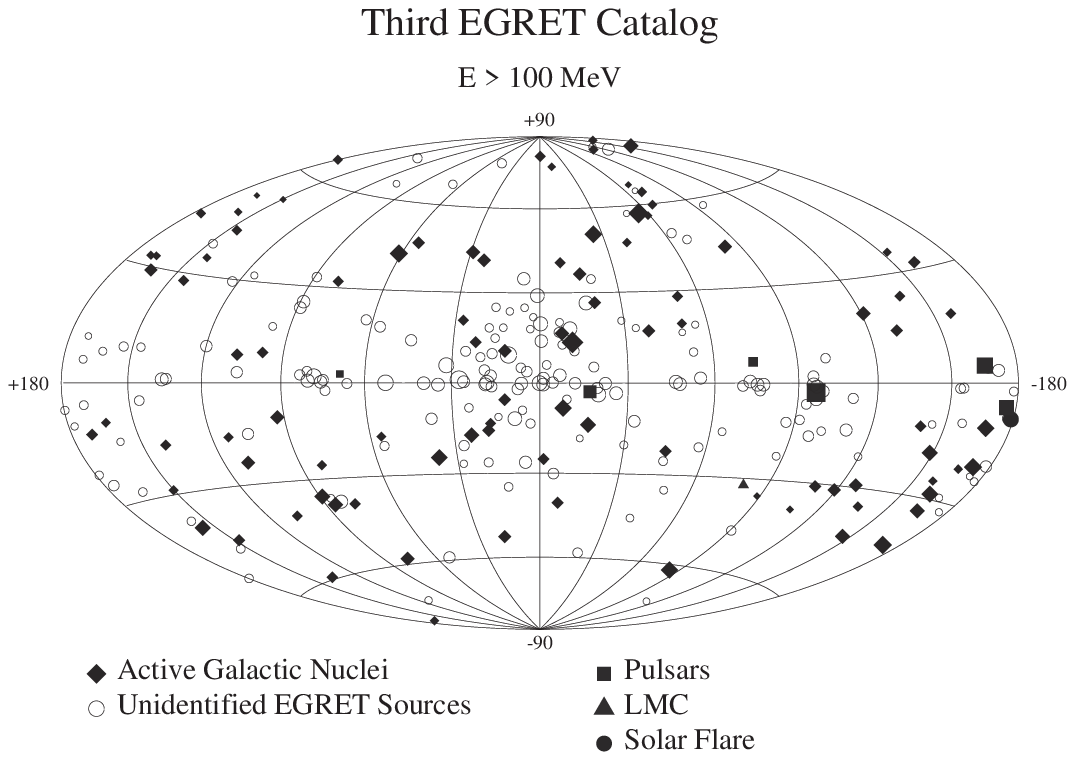, width=1.0\hsize}
\caption{Gamma-ray point sources in the Third EGRET catalog}
\end{figure}
With the pecularities of the construction of the Third EGRET catalog in mind,
nevertheless the longitudinal and latitudinal characteristics of the sample
can be sketched. Fig. 3 shows the complete 3EG catalog sample (outlined) and a selection 
made of sources fulfilling a common (TS)$^{1/2}> 5$ acceptance criterion on the basis of 
the summed map from CGRO observation cycles 1 to 4 only (shaded). As can easily be seen, 
the distribution of unidentified sources represents a distinct Galactic population, but 
also shows some spreading at mid-latitudes on top of a rather 
flat component present at all latitudes.\\
\begin{figure}[ht]
\epsfig{file=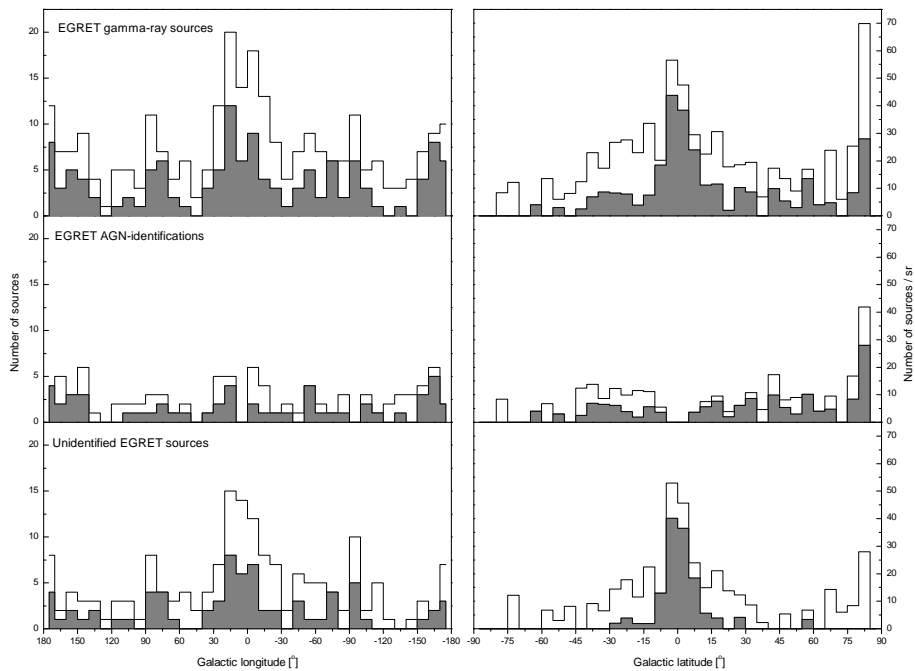, width=1.0\hsize}
\caption{Longitudinal and latitudinal distributions of the 3EG catalog sources. 
See text for details.}
\end{figure}  
Additional compilations of high-energy sources have been made, i.e. for GeV sources 
(Lamb \& Macomb 1997, Reimer et al. 1997). The obvious reason to compile a catalog of 
a higher energy threshold is the significantly reduced Galactic diffuse emission 
component (and therefore surpression of systematic uncertainties originating from it 
during source determination procedures) in conjunction with a narrower instrumental PSF. 
These advantages are offset by the reduced photon flux at higher energies and therefore 
a loss in photon statistics.
However, in cases of bright sources, hard photon spectral indices or regions suffering 
from source confusion, the trade between limited statistics and better angular resolution
often leads to significantly narrower error contours. Figure 4 compares the error contours 
of the unidentified high-latitude source 3EG J2020+4014 ($\gamma$ Cyg) at energies above 
100 MeV and above 1 GeV, respectively.
\begin{figure}[ht]
\epsfig{file=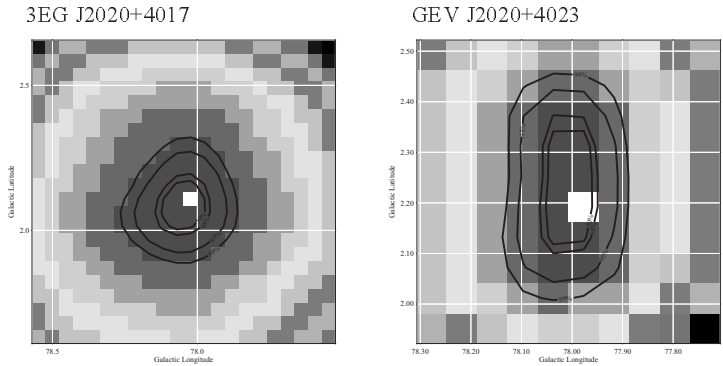, width=0.8\hsize}
\caption{Gamma-ray emission from $\gamma$ Cygni at E $> 100$ MeV and E $> 1$ GeV, respectively.
Note the different scaling of the figures. The visible pixelation of 0.05$^\circ$ by 0.05$^\circ$ 
is the same in both images. The best position is similar, but the uncertainties are smaller at 
higher energies, dedicating the GeV-image for counterpart studies.}
\end{figure}
A proper way to treat error contours in a mathematical-statistical way (i.e. for 
counterpart propability tests) has been found in elliptical fits to source contours. 
Narrower error contours to a gamma-ray source do not automatically imply a more exact
point source location in the coordinates. Comparing the catalog source positions determined
at different energety thresholds with precise coordinates of astronomical objects could only be 
performed if an identification has been established, i.e. from observations at other wavelengths. 
For the high-energy gamma-ray sources this can be accomplished using pulsars (PSRs) and 
active galactic nuclei (AGN). Figure 5 compares the gamma-ray source locations from the 
3EG catalog (Hartman et al. 1999) and the GeV-catalog (Lamb \& Macomb 1997) with the radio 
positions of these objects, providing estimates of the precision with which
EGRET typically determines gamma-ray point source coordinates.
\begin{figure}[ht]
\epsfig{file=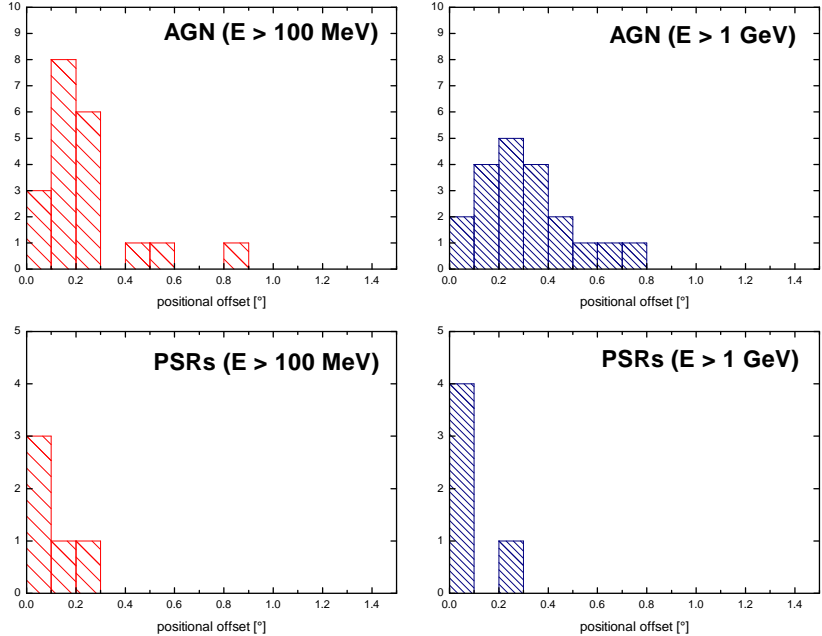, width=0.9\hsize}
\caption{Source location offsets comparing the gamma-ray coordinates of identified objects with 
precise coordinates from optical or radio counterparts. The GeV positions are not significantly
more precise than those determined for E $> 100$ MeV.}
\end{figure}

\section{EGRET Source Detectibility and Consequences}

The significance s for a detection of an isolated point source with EGRET (E$ > 100$ MeV)
is adequately represented by
\begin{equation}
s \propto f \sqrt{\frac{e}{bg}},
\end{equation}
where f is the flux, e the exposure and bg the intensity of the diffuse gamma-ray emission 
at the region of the source (Mattox et al. 1996). In order to simulate a corresponding picture
of the gamma-ray sky as given in the 3EG catalog, all three observables needs to be looked
at in detail. 

\begin{itemize}
\item
Exposure: The observational history of the EGRET instrument is highly non--uniform and so is the 
exposure. EGRET pointings have a typical field--of--view of $40^\circ$, however for most applications 
a $30^\circ$ cut is recommended as applied, for instance, in the 3EG catalog. If population studies 
compare with the EGRET source catalog, the exposure needs to be determined for each individual source. 
As mentioned above, catalog sources are not consistently included on the basis of equal exposure time. 
Therefore, exposure time as well as the corresponding number of sources matching the detection criterion 
at any considered time interval should be looked at.

\item
Diffuse gamma-ray emission: The detectibilty for EGRET sources also depends on the 
diffuse gamma-ray background in the source region. To compensate for these
nonuniformities, several ways are applicable. The diffuse gamma-ray emission model 
constructed from HI and CO distributions (Hunter et al. 1997) is available in the standard
EGRET energy intervals. Alternatively, a point--source--removed intensity map could be used
as has been done by Strong, Moskalenko \& Reimer 2000. Although both approaches are different
in the scientific content, they provide a good measure of the characteristics of the diffuse
gamma-ray emission for the purpose of use in population studies.

\item
Flux: Monte-carlo based population simulations are required to reproduce the actually observed 
log N--log S distribution of gamma-ray sources, globally as well as locally. As noticed by 
Gehrels et al. 2000, the log N--log S distribution for unidentified sources close to the 
Galactic plane differs from the one obtained at high Galactic latitudes, although 
partly as a result of the nonuniform detectibility function of EGRET itself. Nevertheless, the 
log N--log S of identified gamma-ray sources is distinctly different, i.e. for AGN (\"Ozel \& Thompson 1996).
Estimates of the fraction of unresolved point sources will come to play when concluding on the basis of
simulations involving large numbers of sources as expected in the GLAST-era.
\end{itemize}
  
A rather simplified approch to account for the EGRET detectibilty function can be made by  
determining upper limits for a grid on the sky. These upper limits needs to be determined at
comparable statistical significance. An example is given in the 3EG catalog, using the
summed exposures of CGRO cycles 1, 2, 3, and 4. Note that the catalog acceptance criterion
and 95\% confidence upper limits leave room for excesses in the test statistics inbetween,
which needs to be accounted for in simulations. The full composition of the 3EG catalog might
be investigated by following this scheme through each of the individual viewing periods in order 
to account for transient sources. Lastly, upper limits near bright catalog sources are expected to 
be underestimated due to the width of the EGRET point spread function. 

\begin{figure}[ht]
\epsfig{file=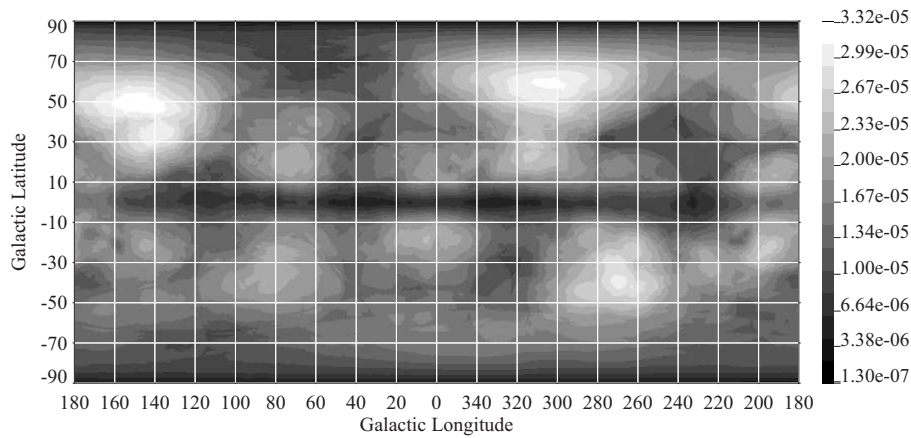, width=1.0\hsize}
\caption{Two-dimensional detectibility function for gamma-ray sources above 100 MeV and
comparable flux. This map corresponds only to the analysis of the summed EGRET cycle 1 to 4 
observations, as listet as P1234 in the 3EG catalog (units of cm$^2$ s).}
\end{figure}

Figure 6 shows a detectibility map for equally bright sources, determined for EGRET observations 
from CGRO cycles 1, 2, 3, and 4 (E$ > 100$ MeV). Note the features close to the Galactic plane,
where the low detectibility directly in the plane increasingly becomes compensated from the 
exposure. However, the highly nonuniform character of EGRETs detectibility function is easy to
recognize. On a $0.5^\circ$ x $0.5^\circ$ grid, the value of the detectibilty function for sources 
of comparable flux differs by more than a factor of ten!  

\section{Variability of Gamma-Ray Sources}

Gamma-ray source variability is even more difficult to quantify than test statistics excesses,
nonuniform detection thresholds and absolute coordinates for gamma-ray sources. At first glance,
one needs to look at the spark chamber efficiency of the EGRET instrument. As discribed in 
Esposito et al. 1999, the spark chamber efficiency is strongly time-dependent. For any meaningful 
determination of gamma-ray source fluxes, the response has to be normalized. Scale-factors have been 
constructed for a given energy and time interval, primarily by comparing the level of the ever present 
diffuse emission component. Neither the determination of the inital response nor the scaling to a nominal 
value is perfect. Although we think that the response of the spark chamber should underly a rather smooth 
degradation between major events impacting its performance (gas refills), some individual viewing periods 
appear to be slightly off the generally smooth normalization scheme. However, the individual normalization 
factors are accessible for each observation period and the individual case could be well investigated beyond 
the flux value listed in a source catalog.  
Figure 7 gives an impression of the spark chamber efficiency versus time, before and after the 
normalization has been applied.\\

\begin{figure}[ht]
\epsfig{file=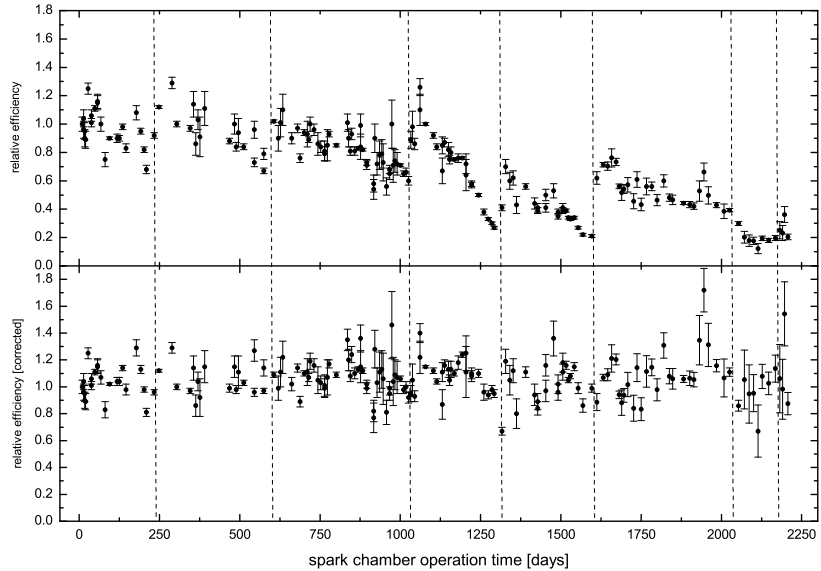, width=0.9\hsize}
\caption{EGRET efficiency as function of observation time, Upper panel: before normalization,
lower panel: after normalization has been applied.}
\end{figure}

So far, variablity studies have been performed largely on the basis of entries in EGRET source catalogs.
The first systematic study by McLaughlin et al. 1996 made use of individual viewing periods from the 
2EG catalog (Thompson et al. 1995), and has been expanded online to include CGRO cycle 3 observations.
It has been used also by Wallace et al. 2000 to study short-term time variability on the basis
of sub-viewing periods. The method is effectively a measure of inconsistency of the gamma-ray
data with the assumption of a constant source flux (for details see McLaughlin et al. 1996).\\
A different approach to properly quantify flux variability has been carried out by Tompkins 1999,
consitently computed for the 3EG catalog sources and source fluxes. A method has been
introduced to obtain a fractional variability measure by calculating the likelihood for obtaining
any source flux and compare to the actual observation (for details see Tompkins 1999).\\
Following a method successfully applied to radio data (Romero, Combi \& Colomb 1994), Zhang et al. 2000
and Torres et al. 2000 determined a gamma-ray flux variability measure by means of a weighted fluctuation index
with pulsars as ``standard candle". Unfortunately, both teams use different subsets of gamma-ray sources
in the variability study which introduces problems to directly compare the results of the different methods.\\
Although being a selection of steady sources by comparing detection significances at different observational 
timescales only (individual and combined viewing periods), Gehrels et al. 2000 effectively obtained a qualitative 
measure of source variabilty. This source ensemble is selected against one-time flaring sources 
(transients) and variable but dim sources close to the detection threshold of the EGRET instrument.

Each method attempted to distinguish gamma-ray sources by means of quantifying the degree of flux
variability. At present, different classes of gamma-ray emitters are most clearly distinguishable 
in terms of variabilty by applying the method developed by Tompkins. Figure 8 gives the sketch of
the separation achieved on the basis of the variabilty criterion $\tau$ (as the inverse fraction 
of the average and the standard deviation). Among the unidentified sources, there is a tendency
that sources at low Galactic latitudes are less variable than at high Galactic latitudes. It is
striking that the variability distribution of unidentified sources as a whole is distincly different 
than the one of pulsars and active galactic nuclei.\\

\begin{figure}[ht]
\epsfig{file=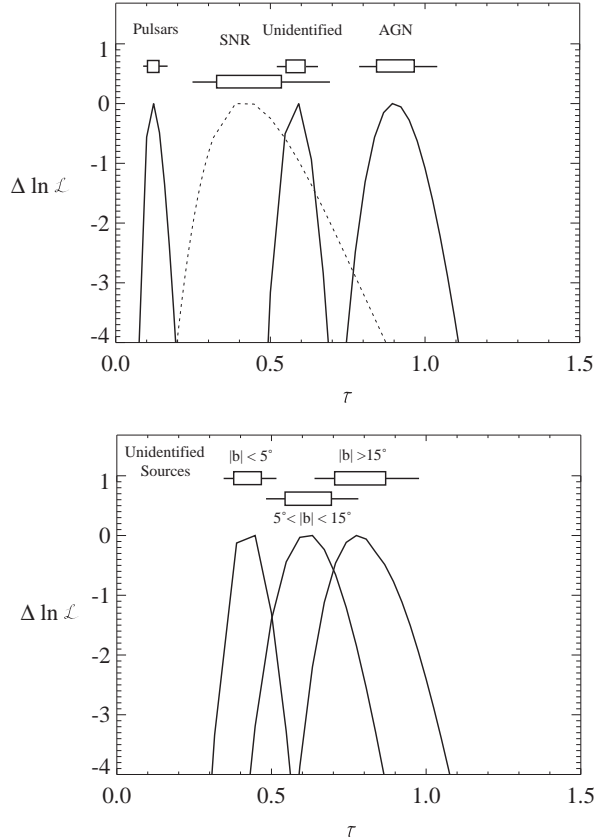, width=0.65\hsize}
\caption{Variabilty index for different source populations. From: [Tompkins 1999]}
\end{figure}

When comparing the results of the different methods against each other, the level 
of consistency among the results of the variability studies is incredibly low. Unfortunately, 
each study has been performed using different source ensembles, hampering efforts to trace 
the origin of such discrepancies. Figure 9 compares, for instance, results 
from the variability study by Tompkins with the one by McLaughlin, using sources common to two samples.
Note that the difference of the number of viewing periods included in the study by Tompkins and 
by McLaughlin could not account for the spread in the variability results.\\

\begin{figure}[ht]
\epsfig{file=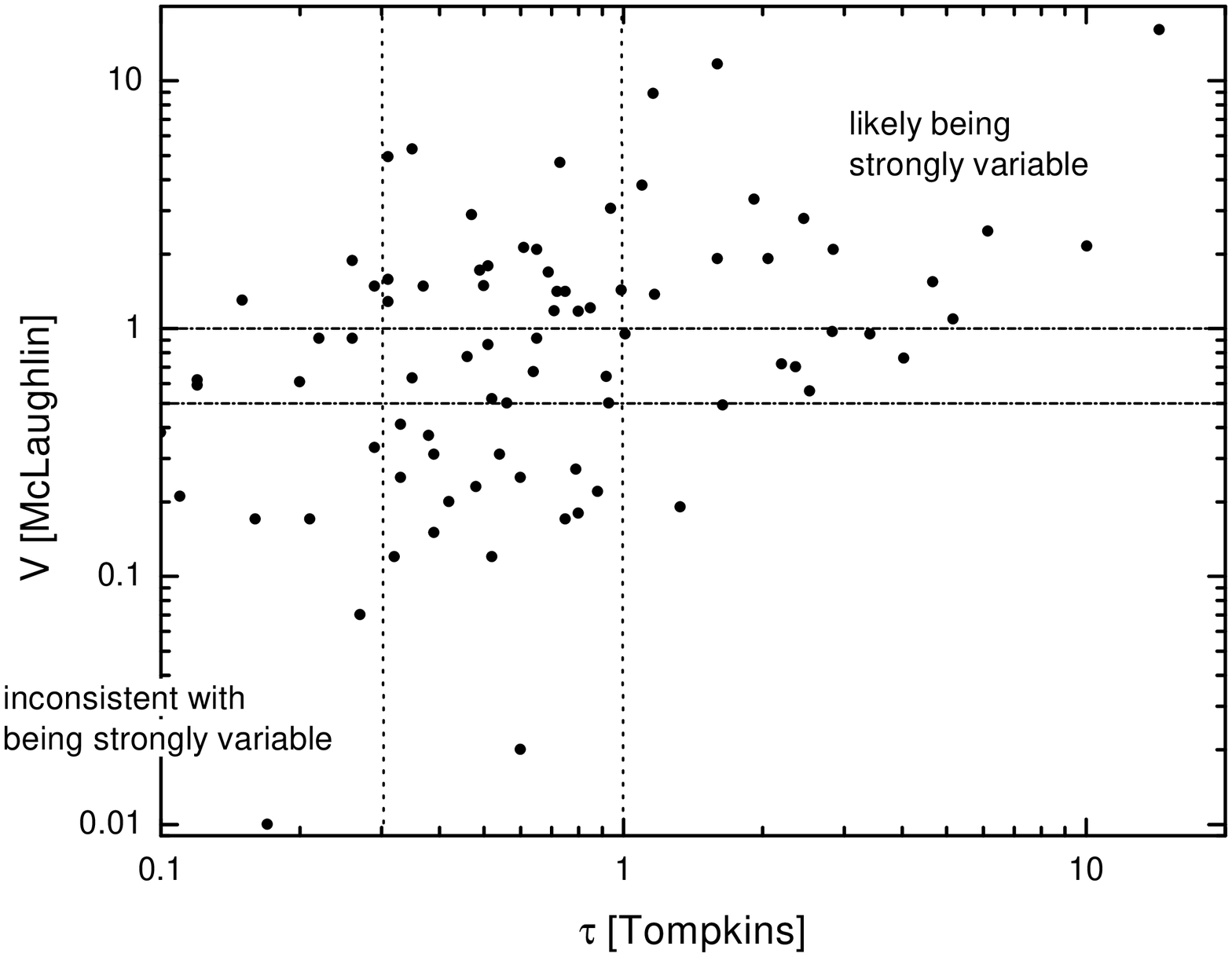, width=0.8\hsize}
\caption{Comparison of the results from different variability studies, here
given for McLaughlin et al. 1996 and Tompkins 1999. The vertical and horizontal 
grid represent the suggested boundaries by the autors for low variabilty, uncertain, and 
high variability characteristics.}
\end{figure}

In consequence, not only the flux uncertainty of a particular source in a given time interval 
needs to be accounted for, also the uncertainties of the normalization procedure. Studies aimed 
to compute a common measure of gamma-ray flux variation should be aware of the level of underlying
systematics before drawing conclusions. Variabilty examined up to the very detail for an individual
source might stand here as an example, as recently revised for the high-latitude unidentified
source 3EG J1835+5918 (Reimer et al. 2001). For many years believed to be a variable source of gamma-ray 
emission, only the combination of its expanded observational history, appropriate cuts to establish comparable 
quality in the data to be compared and an understanding of the systematics revealed that this source 
actually is compatible being a non-variable source, in this case with direct implications of its likely 
identification.
      
\section{Spectral Characteristics of Gamma-Ray Sources} 

The efficency of the EGRET spark chamber is not only a function of time, but also of
the energy. Figure 10 shows the spark chamber efficiency as function of time for the
ten energy intervals typically chosen for determining the spectrum of an
EGRET detected gamma-ray source. For clarification, the normalization functions 
obtained by fitting the data (compare with upper panel Fig.7) are given here. Consequences 
beyond the systematics as already mentioned for gamma-ray source fluxes are expected.

\begin{figure}[ht]
\epsfig{file=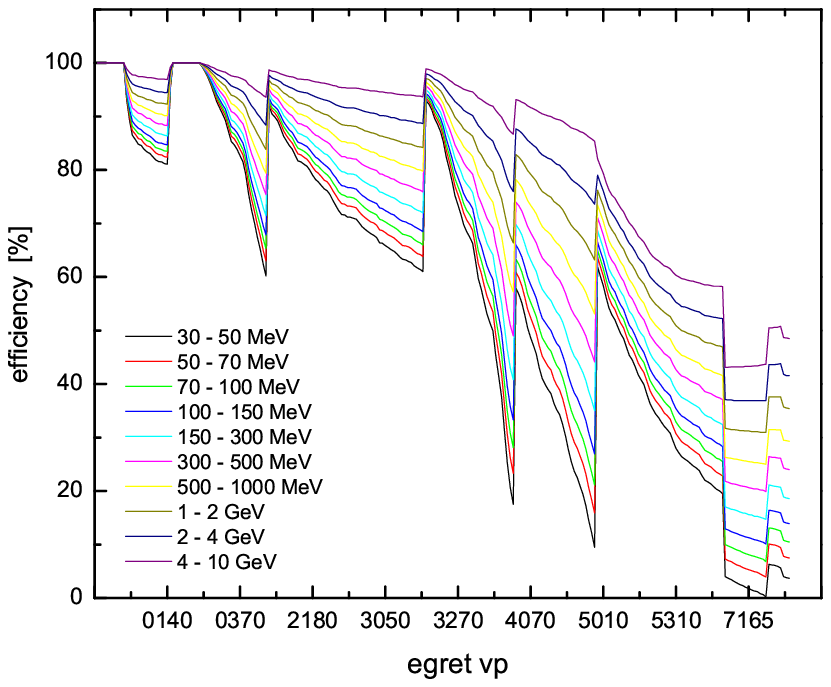, width=1.0\hsize}
\caption{EGRET efficiency as function of observation time and energy. The ten subsequent
energy bands are sketched which are used to determine source spectra. For clearer view,
here the correction functions are plotted only.}
\end{figure}

The Third EGRET catalog lists for the majority of the sources the photon spectral index in 
F(E) $\sim$ E$^{-\gamma}$, consistently determined for the combined observations from CGRO 
observation cycles 1 to 4. In various cases the P1234 sum is not the most significant detection 
and a meaningful or even better spectrum could be determined in an individual viewing period.
For the EGRET detected AGN a spectral study on the level of individual viewing periods has
been performed (if sufficient counts have been recorded) by Mukherjee et al. 1997. Fierro et al.
1997 also published phase-resolved spectra for the brightest gamma-ray pulsars. At present,
individual EGRET source spectra are investigated beyond a single power-law model fit (Bertsch et al. 2000). 
Additionally, the spectral index could be used in order to conclude on
spectral variability. Such information goes beyond flux variability and an individual
power-law spectral index only, but has not often be used in studies of individual as well 
as population studies so far.\\
\begin{figure}[ht]
\epsfig{file=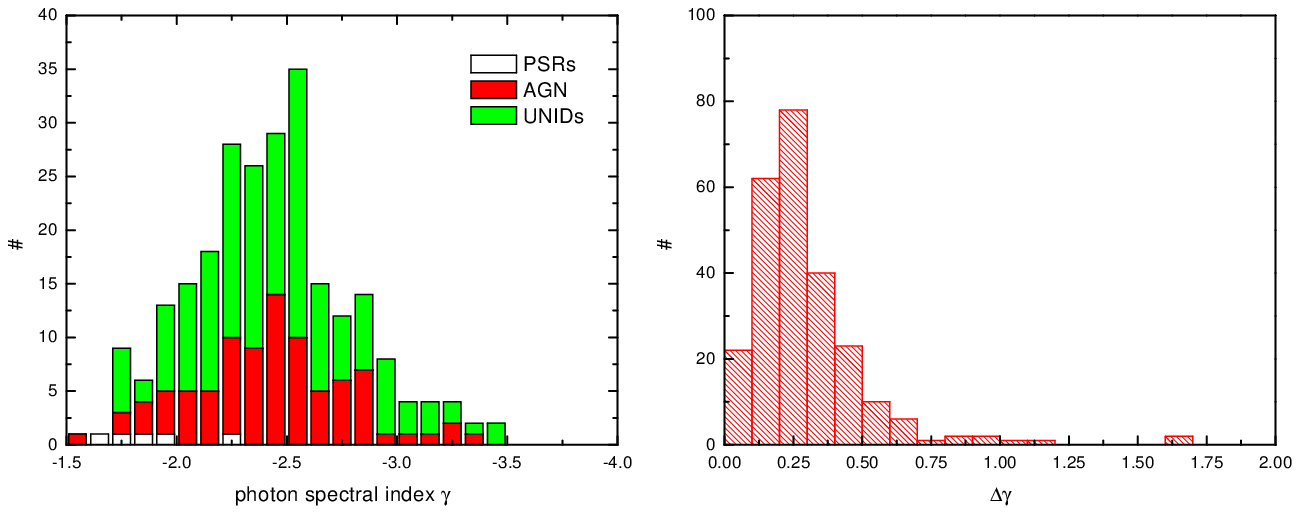, width=1.0\hsize}
\caption{Distribution of the power--law spectral index for AGN, PSRs and unidentified 
EGRET sources and its uncertainty. It gives a vague impression about the difficulty 
to distinguish unidentified gamma-ray sources by its spectral characteristics only.}
\end{figure}
The spectral characteristics of individual gamma-ray sources have been used in a similar way 
to attempt a distinction between source classes in population studies as variabilty, exclusivly
on the basis of the photon spectral indices given in the 3EG catalog. Generally ignored by only picking up
the numerical value of the spectral index when drawing conclusions, the significant uncertainties 
in the spectral index, especially for dim sources, put conclusions of distinct spectral characteristics
rather into perspective. At present, conservative conclusions could only be drawn from the hardness of 
the spectrum of gamma-ray pulsars and the indication of a cut-off at GeV-energies. AGN seem to have 
a different spectrum when observed at high activity states (outbursts/flares) compared to their 
average spectra, however this needs to be investigated further. The vast majority of the unidentified 
EGRET sources are currently not distinguishable from the identified gamma-ray source populations by
spectral characteristics only. Due to the wide spread found among the spectal index, and accounting 
for the uncertainties in the power-law spectral fits as well as the systematic bias towards finding
hard spectrum gamma-ray sources significantly easier to discriminate against the diffuse
emission component than soft spectra sources, it appears that viable conclusions might
arise rather from the detailed spectrum of an individual source than from populations studies. 
Merck et al. 1996 conducted a survey for pulsar-like characteristics among the sources near
the Galactic plane. At present, three of the sources suggested therein support the validity of 
this approach (PSR B1046-58/3EG J1048-5840, RX J2020.2+4026/3EG J2020+4017, SAX J0635+0533/3EG J0634+0521).

\section{Signatures from spatial, temporal and spectral properties} 

Seeking signatures from the spatial, temporal and spectral characteristics of
the unidentified EGRET sources is the subject of population studies. At present,
the ensemble of persistent (Grenier 1999) or steady sources (Gehrels et al. 2000) 
appears to be the most promising in order to give hints on the nature of unidentified sources. 
Here I only sketch the general associations between spatial, temporal, and spectral
properties for the unidentified EGRET sources and address the most prominent features 
or lack of features, respectively.

\begin{figure}[ht]
\epsfig{file=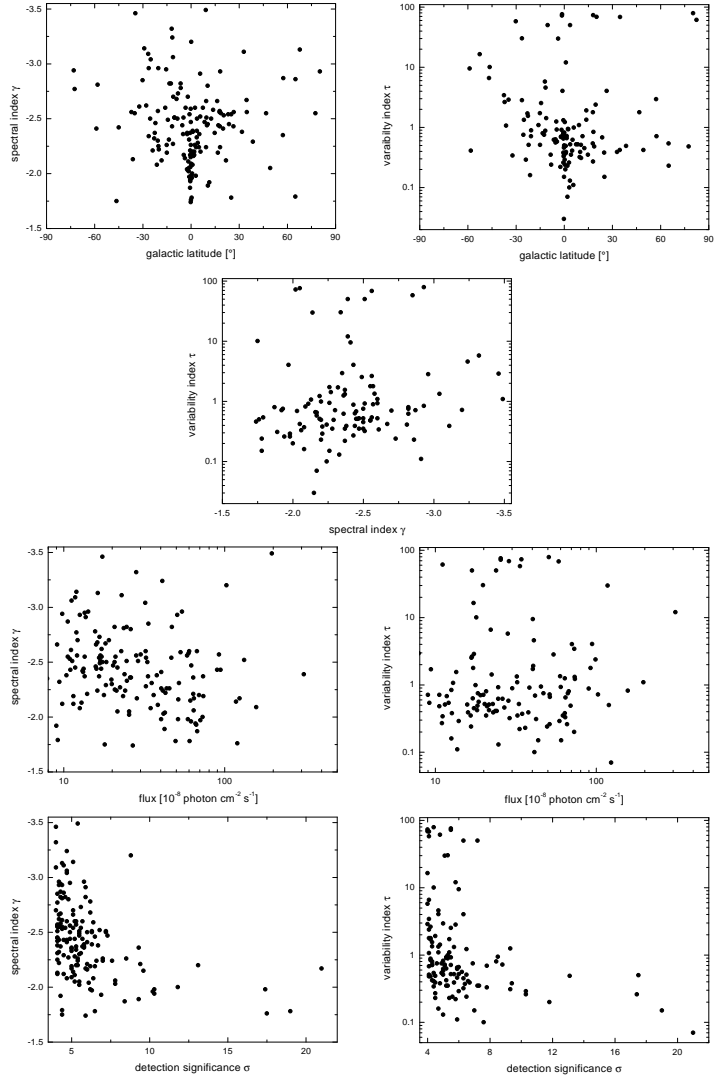, width=0.8\hsize}
\caption{Associations between the observables determined to characterize the gamma-ray
properties of unidentified EGRET sources. See text for details.}
\end{figure}

In Fig. 12a, the tendency to find hard spectrum sources predominantely close to the Galactic
plane is apparent. However, this is at least partly due to the inability to discriminate soft
spectrum sources near the plane against the dominant diffuse gamma-ray emission component. 
Fig. 12b gives the impression that variability is a common phenomenon for unidentified sources 
at all Galactic latitudes, although a clustering of more steady sources towards the Galactic
plane is indicated, at least for the sources with extremely low probability to exhibit strong
flux variability. Fig. 12c shows the correlation between spectral and variability index. Only 
a rough tendency is indicated, that the softer sources are also characterized by higher 
variability. Counterexamples could be found for both extremes: hard spectrum sources exhibiting
a high degree of flux variability and soft sources consistent with being nonvariable. Fig. 12d
shows the flux distribution of unidentified sources. The few extremely high fluxes are the signatures 
of transient sources, arising from hard as well as soft spectrum emitters.
Fig. 12e shows the same flux distribution against the variability index. The flaring or transient
behaviour of unidentified sources is not in direct relation to either extrem flux values or strong 
source variability. This seems rather curious, but becomes obvious when comparing with 
a similar arrangement of these quantities for the EGRET detected AGN. Fig. 12f and 12g show 
spectral and variability index as function of the source detection significance. High detection 
significances could be found predominantly for hard spectrum and low variable sources, giving the 
combination of both the rather distinct feature as apparent in the selections called "steady" 
(Gehrels et al. 2000) or "persistent" (Grenier et al. 1999).\\ 
However, nearly all signatures have to be put into perspective when supplemented by the appropriate
uncertainties. The less apparent correlations are basically at the $1 \sigma$ level in the 
uncertainties of the observables themselves. Only carefully chosen combinations between the observables 
still reveal higher degrees of confidence in correlations among the observable parameter 
of the unidentified EGRET sources.

\section{Conclusions}

At present, we still fail to identify the nature of the majority of gamma-ray sources
on the sky. However, the various observables at hand help to characterize individual 
sources rather well. For applications beyond individual sources (population studies,
selected source ensembles) the degree of systematic biases and individual uncertainties
needs to accounted for. Above all, in most cases gamma-ray astronomy still suffers from
statistical limitations. Applying cuts is a valid procedure only if the particular cut
is well understood in all its consequences for the data space. Conclusions drawn from
subsets have to address its implication for the residual sources also. Generally, 
selections have to be made to avoid additional non--uniformities as already present 
in the 3EG source catalog. It needs to be investigated, at which level
the known systematics and biases might put existing population studies in perspective. 
Efforts should be made to obtain more unbiased subsets instead of picking up any and
each of the catalog listed gamma-ray sources. A compensation against the various biases by 
performing appropriate corrections is a requirement for comprehensive population studies.\\ 
The EGRET data will remain unique in gamma-ray astronomy until follow-up missions, 
in particular GLAST, will clarify a lot concerning the identity of individal gamma-ray 
sources and hypotheses of the composition of unidentified EGRET sources in the collective. 
Until then we have to work out the open questions on the basis of already acquired 
data. The tremendous potential offered by the nine years of EGRET data should not be neglected.

\begin{acknowledgments}
The author wish to thank Alberto Carrami\~nana and the staff of 
INAOE for the kind hospitality to make this workshop happened
in a way as successfully and memorable as it apparently became.
Also, I like to thank NRC for travel support. 
\end{acknowledgments}

\end{document}